\documentclass[aps,psfig,epsfig,preprint] {revtex4}
\usepackage{epsfig}
\usepackage{color}
\usepackage{bm}
\usepackage{subfigure}

\begin{document}
\draft
\title{
{\bf Understanding the Charged Meson Z(4430) }}

\author{Gui-Jun Ding}

\affiliation{\centerline{Department of Modern
Physics,}\centerline{University of Science and Technology of
China,Hefei, Anhui 230026, China}} 

\begin{abstract}

The difference between Z(4430) as a $D^{*}D_1$ molecule and a
tetraquark state and how to distinguish between them are discussed.
We construct an effective Lagrangian with $D^{*}D_1$ contact
interactions constrained by the heavy quark symmetry and chiral
symmetry to study Z(4430). We find that if Z(4430) is a $D^{*}D_1$
molecule state, there should be a $B^{*}B_1$ bound state as well,
and it mass is about 11048.6 MeV.

\vskip 0.5cm


\end{abstract}
\maketitle
\section{introduction}

Recently the Belle Collaboration has reported a new state Z(4430) in
the $\pi^{+}\psi'$ invariant mass spectrum in $B\rightarrow
K\pi^{+}\psi'$ with statistical significance greater than
$7\sigma$\cite{2007wga}. The Breit Wigner fit for this resonance
yields the peak mass ${{ M_{Z}}=4433\pm4({\rm stat})\pm1({\rm syst})
{\rm MeV}}$ and the width ${\rm
\Gamma=44^{+17}_{-13}(stat)^{+30}_{-11}(syst) MeV}$. The product
branching fraction is determined to be $\mathcal{B}(B\rightarrow
KZ(4430))\times\mathcal{B}(Z(4430)\rightarrow\pi^{+}\psi')=(4.1\pm1.0({\rm
stat})\pm1.3({\rm syst}))\times10^{-5}$. Differing from other hidden
charmonium-like states such as X(3872) and Y(4260) etc, Z(4430) is a
positively charged state, therefore it must not be a conventional
$c\bar{c}$ state. It would be an exotic state beyond the naive
quenched quark model, if it is confirmed by the further experiments.

Some theoretical studies have been carried out to understand the
structure and properties of this interesting state. Because its mass
is so close to the threshold of ${D^{*}\overline{D}_1(2420)}$,
Rosner suggested that Z(4430) is a S-wave threshold
effect\cite{Rosner:2007mu}. Meng and Chao proposed that Z(4430) is a
S-wave ${D^{*}D_1}$(or ${D^{*}D_1'}$) resonance, re-scattering
mechanism has been suggested to explain the absence of the signal in
$\pi^{+}J/\psi$ for properly chosen parameters\cite{Meng:2007fu}.
The mass of Z(4430) as a ${\rm J^{P}=0^{-}}$ ${D^{*}D_1}$ molecule
was calculated from the QCD sum rule\cite{Lee:2007gs}. A dynamical
study of whether Z(4430) could be a S-wave molecular state of
${D^{*}D_1}$(or ${D^{*}D_1'}$) has been performed\cite{Liu:2007bf},
where the authors assumed that the long distance one pion exchange
dominates. They found that the attraction from the one pion exchange
potential alone is not strong enough to form a bound ${D^{*}D_1}$(or
${D^{*}D_1'}$) molecular state. Short range force maybe plays an
important role in the dynamics of Z(4430).

Maiani et al suggested that Z(4430) is a diquark-antidiquark state
with flavor ${[cu][\overline{c}\overline{d}]}$, it is the radial
excitations of ${\rm X}^{+}_{u\overline{d}}(1^{+-};1{\rm S})$ with
mass about 3880 MeV, which mainly decays into $J/\psi\pi^{+}$ and
$\eta_c(1{\rm S})\rho^{+}$\cite{Maiani:2007wz}. Tetraquark
interpretation is also suggested based on the QCD-string
model\cite{Gershtein:2007vi}. Other theoretical interpretations such
as baryonium\cite{Qiao:2007ce} and threshold cusp effect are put
forward\cite{Bugg:2007vp}. The mass and the production of the bottom
analog of Z(4430) have been studied as
well\cite{Cheung:2007wf,Li:2007bh}.

Just as X(3872) may be a weakly bound state of
${DD^{*}}$\cite{Tornqvist:2004qy,Close:2003sg,Wong:2003xk,Swanson:2003tb},
the closeness of Z(4430) to the ${D^{*}{D}_1(2420)}$ threshold
strongly suggests that Z(4430) could be a weakly bound
${D^{*}{D}_1(2420)}$ molecular state. This is a old and very
interesting idea which has been applied to a variety of mesons with
unusual characteristics such as the $\psi(4040)$\cite{russa,De
Rujula:1976qd} and $f_0(980)$\cite{Weinstein:1990gu}. Although the
mass of Z(4430) is also close to the threshold of ${
D^{*}(2010)D'_1(2430)}$, ${D'_1(2430)}$ is very broad, therefore it
decays so quickly that it isn't possible to form a ${ D^{*}D'_1}$
molecular state. The component of
${D^{*}(2010)\overline{D}'_1(2430)}$(or
${\overline{D}^{\,*}(2010){D}'_1(2430)}$) could be neglected in the
molecular state interpretation for Z(4430).

If Z(4430) is a weakly bound molecular state, it plays the role of
deuteron in the meson antimeson interactions, which is sometimes
called deuson\cite{Tornqvist:1991ks}. So we can apply the methods
developed for the description of deuteron to
Z(4430)\cite{Weinberg:1990rz,Ordonez:1992xp,van
Kolck:1998bw,Kaplan:1998tg}. We will use an effective field theory
to describe Z(4430), which is similar to the pionless effective
theory of shallow nuclear bound
state\cite{Weinberg:1990rz,Ordonez:1992xp,van Kolck:1998bw}. Since
the binding energy of Z(4430) is small, the size of this bound state
is quite large. Consequently the particular details of the
interactions between the heavy mesons and antimesons are irrelevant
to the description of the molecule state, and we can use the
effective lagrangian with four-meson interactions consistent with
both the heavy quark symmetry and chiral symmetry to study this
system.

The paper is organized as follows. We discuss the crucial signals
which can distinguish between the molecule and tetraquark
interpretation in Sec.II. In Sec.III we construct the effective
Lagrangian consistent with heavy quark symmetry and chiral symmetry,
The binding of Z(4430) is studied by considering the transition
amplitude for ${\rm Z(4430)\rightarrow Z(4430)}$. Taking into
account the scaling of the effective coupling constant, we predict
the mass of the bottom analog of Z(4430). A summary of our results
is given in Sec. IV.

\section{Z(4430): Molecule or tetraquark? }

Since the S-wave inter-hadron forces are strongest, it is natural to
expect that ${D^{*}{D}_1}$ is in relative S-wave, then the quantum
number ${\rm J^{P}}$ of Z(4430) can be $0^{-}$, $1^{-}$ and $2^{-}$.
For the $2^{-}$ assignment, its production in ${B\rightarrow
Z(4430)K}$ is strongly suppressed by the small phase space. the
$1^{-}$ state has a larger mass and $0^{-}$ state should be more
stable as suggested by the authors in\cite{Lee:2007gs}. So we assume
$Z(4430)$ as a ${D^{*}D_1}$ molecule with ${\rm J^{P}=0^{-}}$ in
this work. Since Z(4430) was reconstructed in the $\pi^{+}\psi'$
final state, from isospin and $G$-parity conservation, we learn
Z(4430) is a isovector state with positive $G$-parity. Under
$G-$parity transformation,
$\overline{D}^{*0}(\overline{D}^{0}_1)\rightarrow D^{*+}(D^{+}_1)$
and
$D^{*+}(D^{+}_{1})\rightarrow-\overline{D}^{*0}(-\overline{D}^{0}_{1})$,
therefore the flavor wavefunction of Z(4430) is
\begin{equation}
\label{1}{|Z(4430)\rangle=\frac{1}{\sqrt{2}}(|D^{*+}\overline{D}^{\,0}_1\rangle-|D^+_1\overline{D}^{\,*0}\rangle)}
\end{equation}

In \cite{Maiani:2004vq}, Maiani et al. predicted two $1^{+-}$
states, and their masses are approximately 3754 MeV and 3882 MeV
respectively. They identified Z(4430) with the first radial
excitation of the higher $1^{+-}$ state, then the radial excitation
of the lower $1^{+-}$ state with mass about 4344 MeV should be
observed in the $\psi^{'}\pi^{+}$ final state as well. Searching
this state at Belle or Babar is an important test of the structure
of Z(4430).

The difference between the molecular state and the tetraquark
interpretation is obvious. In the tetraquark picture, ${\rm J^{P}}$
of Z(4430) is $1^{+}$ which is different from $0^{-}$, $1^{-}$ or
$2^{-}$ in the S-wave ${D^{*}D_1}$ molecular state case. For the
molecule interpretation, the leading source of decay is
dissociation, to good approximation dissociation will proceed via
the free space decay of the constituent mesons. Since ${D_1}$
dominantly decays into ${D^{*}\pi}$, ${D^{*}D^{*}\pi}$ should be the
main decay mode for Z(4430) as a ${D^{*}D_1}$ molecule. While the
decay of Z(4430) could proceed through the "fall apart" decay
mechanism in the tetraquark picture, it can decay into ${DD^{*}}$
and ${D^{*}D^{*}}$ in both S-wave and D-wave besides the {\rm
$J/\psi\pi$, $J/\psi\rho$, $\eta_c(1{\rm S})\rho$ and $\psi(2{\rm
S})\pi$} final states, however, it can not decay into ${DD}$ due to
its unnatural spin-parity. So whether the three body mode
${D^{*}D^{*}\pi}$ has considerable branch ratios or the two body
decay modes ${DD^{*}}$, ${D^{*}D^{*}}$, {\rm $J/\psi\pi$,
$J/\psi\rho$, $\eta_c(1{\rm S})\rho$ and $\psi(2{\rm S})\pi$} is
another important test of the nature of Z(4430).

\section{Z(4430) as a ${D^{*}D_1}$ molecule from the effective field theory}

The general effective Lagrangian required to describe the
${D^{*}D_1}$ molecule is constrained by both the heavy quark
symmetry and chiral symmetry, it consists of the one-body
interaction terms and the two-body interaction terms
\begin{equation}
\label{2}\mathcal{L}=\mathcal{L}_1+\mathcal{L}_2
\end{equation}
The one-body effective lagrangian ${\mathcal{L}}_1$ which describes
the strong interaction of heavy meson with one heavy
quark(antiquark) is given by the heavy-hadron chiral perturbation
theory\cite{Wise:1992hn,Burdman:1992gh,Yan:1992gz,Falk:1992cx}
\begin{eqnarray}
\nonumber\mathcal{L}_1&&=-i{\rm Tr}[\overline{H}^{\,(Q)}_a(v\cdot
D_{ba}+\frac{D^{2}_{ba}}{2m_{P}})H^{(Q)}_b]+\frac{i}{2}g\,{\rm Tr}[\overline{H}^{\,(Q)}_aH^{(Q)}_{b}\gamma_{\mu}\gamma_5(\xi^{\dagger}\partial^{\mu}\xi-\xi\partial^{\mu}\xi^{\dagger})_{ba}]\\
\nonumber&&+\frac{\lambda_2}{m_{Q}}{\rm
Tr}[\overline{H}^{\,(Q)}_a\sigma^{\mu\nu}H^{(Q)}_{a}\sigma_{\mu\nu}]
+{\rm Tr}[\overline{T}^{\,(Q)\mu}_a(iv\cdot D_{ba}-\delta
m_T\delta_{ba}+\frac{D^{2}_{ba}}{2m_{T}})T^{\,(Q)}_{\mu
b}]\\
\nonumber&&+\frac{i}{2}g''\,{\rm
Tr}[\overline{T}^{\,(Q)\mu}_aT^{(Q)}_{\mu
b}\gamma_{\nu}\gamma_5(\xi^{\dagger}\partial^{\nu}\xi-\xi\partial^{\nu}\xi^{\dagger})_{ba}]+(H^{(Q)}_a\rightarrow
\overline{H}^{\,(\overline{Q})}_a,\overline{H}^{\,(Q)}_a\rightarrow
{H}^{\,(\overline{Q})}_a,\\
\label{3}&&T^{(Q)\mu}_a\rightarrow
\overline{T}^{\,(\overline{Q})\mu}_a,\overline{T}^{\,(Q)\mu}_a\rightarrow
{T}^{\,(\overline{Q})\mu}_a)+...
\end{eqnarray}
where $H^{(Q)}_a$ and $T^{(Q)\mu}_a$ are the matrix representations
of the heavy mesons, $H^{(\overline{Q})}_a$ and
$T^{(\overline{Q})\mu}_a$ are the matrix representations of the
heavy antimesons, and the ellipsis denotes higher order terms in the
chiral expansion. The covariant derivative
$D^{\mu}_{ab}=\partial^{\mu}\delta_{ab}-\frac{1}{2}(\xi^{\dagger}\partial^{\mu}\xi+\xi\partial^{\mu}\xi^{\dagger})$,
$\delta m_{T}=M_{P_1}-M_{P}$, and the mass difference between
$P^{*}$ and $P$ is $\Delta\equiv
{M_{P^{*}}}-M_{P}=-\frac{8\lambda_2}{m_Q}$. The superfield
multiplets $H^{(Q)}_a$, $T^{(Q)\mu}_a$, $H^{(\overline{Q})}_a$ and
$T^{(\overline{Q})\mu}_a$ are as follows
\begin{eqnarray}
\nonumber H^{(Q)}_a&=&\frac{1+\not
v}{2}[P^{(Q)*\mu}_a\gamma_{\mu}-P^{(Q)}_a\gamma_5]\\
\nonumber T^{(Q)\mu}_a&=&\frac{1+\not
v}{2}\{P^{(Q)*\mu\nu}_{2a}\gamma_{\nu}-\sqrt{\frac{3}{2}}P^{(Q)\nu}_{1a}\gamma_5[g^{\mu}_{\nu}-\frac{1}{3}\gamma_{\nu}(\gamma^{\mu}-v^{\mu})]\}\\
\nonumber
H^{(\overline{Q})}_a&=&[P^{(\overline{Q})*\mu}_a\gamma_{\mu}-P^{(\overline{Q})}_a\gamma_5]\frac{1-\not
v }{2}\\
\label{4}
T^{(\overline{Q})\mu}_a&=&\{P^{(\overline{Q})*\mu\nu}_{2a}\gamma_{\nu}-\sqrt{\frac{3}{2}}P^{(\overline{Q})\nu}_{1a}\gamma_5[g^{\mu}_{\nu}-\frac{1}{3}(\gamma^{\mu}-v^{\mu})\gamma_{\nu}]\}\frac{1-\not
v }{2}
\end{eqnarray}
The pseudogoldstone boson octect is introduced via the exponential
representation
\begin{equation}
\label{5}\xi=\exp(i\mathcal{M}/f_{\pi}),~~~\Sigma=\xi^2
\end{equation}
with $f_{\pi}=132$ MeV and
\begin{equation}
\label{6}\mathcal{M}=\left(\begin{array}{ccc}
\frac{1}{\sqrt{2}}\pi^{0}+\frac{1}{\sqrt{6}}\eta&\pi^{+}&K^{+}\\
\pi^{-}&-\frac{1}{\sqrt{2}}\pi^{0}+\frac{1}{\sqrt{6}}\eta&
K^{0}\\
K^- &\bar{K}^{0}&-\sqrt{\frac{2}{3}}\,\eta
\end{array}\right)
\end{equation}

The two-body effective Lagrangian $\mathcal{L}_2$ which describes
the interactions between the heavy mesons and antimesons is the
local four-boson contact interactions as follows
\begin{eqnarray}
\nonumber&&\mathcal{L}_2=\frac{1}{4}h_1\{{\rm
Tr}[\overline{H}^{\,(Q)}_aH^{(Q)}_a\gamma_{\mu}]{\rm
 Tr}[T^{(\overline{Q})\alpha}_b\overline{T}^{\,(\overline{Q})}_{b\alpha
}\gamma^{\mu}]+{\rm
 Tr}[\overline{T}^{\,(Q)\alpha}_aT^{(Q)}_{a\alpha
}\gamma_{\mu}]{\rm
 Tr}[H^{(\overline{Q})}_b\overline{H}^{\,(\overline{Q})}_b\gamma^{\mu}]\}\\
\label{7}&&+\frac{3}{10}h_2\{{\rm
 Tr}[\overline{H}^{\,(Q)}_aH^{(Q)}_a\gamma_{\mu}\gamma_5]{\rm
 Tr}[T^{(\overline{Q})\alpha}_b\overline{T}^{\,(\overline{Q})}_{b\alpha
}\gamma^{\mu}\gamma_5]+{ \rm
 Tr}[ \overline{T}^{\,(Q)\alpha}_aT^{(Q)}_{a\alpha
}\gamma_{\mu}\gamma_5]{\rm
 Tr}[H^{(\overline{Q})}_b\overline{H}^{\,(\overline{Q})}_b\gamma^{\mu}\gamma_5]\}
\end{eqnarray}
If Z(4430) is indeed a $D^{*}D_1$ molecule, heavy quark symmetry
requires the existence of a ${B^{*}B_1}$ molecular state. To predict
the properties of ${B^{*}B_1}$ molecule from Z(4430), we need to
determine how the coupling constants $h_1$ and $h_2$ scale with the
heavy meson mass $M$. We rescale all energy
$q^{0}\rightarrow\tilde{q}^{0}/M$ and the coordinate $t\rightarrow
M\tilde{t}$ so that the dimensional quantities have the same size(
ie., are measured in unit of the momentum $p$ ). If we demand that
the action is independent of $M$, then since the measure $d^4x\sim
M$, the Lagrangian density $\mathcal{L}\sim 1/M$. The kinetic term
determines that the heavy meson field $H^{(Q)}_a$
($H^{(\overline{Q})}_a$) and $T^{(Q)\mu}_a$
($T^{(\overline{Q})\mu}_a$) scale as $M^{0}$, so the couplings
\begin{equation}
\label{8}h_1\sim h_2\sim 1/M
\end{equation}
The two-body interaction terms relevant to the $P^{*}P_1$ part is
\begin{eqnarray}
\nonumber\mathcal{L}_{2,\,P^{*}P_1}&&=h_1[P^{(Q)*\alpha\dagger}_aP^{(Q)*}_{a\alpha
}P^{(\overline{Q})\beta\dagger}_{1b}P^{(\overline{Q})}_{1b\beta}+P^{(Q)\alpha\dagger}_{1a}P^{(Q)}_{1a\alpha}P^{(\overline{Q})*\beta\dagger}_bP^{(\overline{Q})*}_{b\beta}]\\
\nonumber&&+h_2[P^{(Q)*\alpha\dagger}_aP^{(\overline{Q})}_{1b\alpha}P^{(\overline{Q})\beta\dagger}_{1b}P^{(Q)*}_{a\beta}-P^{(Q)*\alpha\dagger}_aP^{(\overline{Q})\dagger}_{1b\alpha}P^{(Q)*\beta}_aP^{(\overline{Q})}_{1b\beta}\\
\label{9}&&+P^{(Q)\alpha\dagger}_{1a}P^{(\overline{Q})*}_{b\alpha}P^{(\overline{Q})*\beta\dagger}_bP^{(Q)}_{1a\beta}-P^{(Q)\alpha\dagger}_{1a}P^{(\overline{Q})*\dagger}_{b\alpha}P^{(Q)\beta}_{1a}P^{(\overline{Q})*}_{b\beta}]
\end{eqnarray}
Here we concentrate on Z(4430), which we assume to be a $D^{*}D_1$
bound state. Setting $a=2$ and $b=1$ in Eq.(\ref{9}), we obtain the
effective interactions relevant to Z(4430)
\begin{eqnarray}
\nonumber\mathcal{L}_{2,Z}&=&h_1[D^{*+\alpha\dagger}D^{*+}_{\alpha}\overline{D}^{\,0\beta\dagger}_1\overline{D}^{\,0}_{1\beta}+D^{+\alpha\dagger}_1D^{+}_{1\alpha}\overline{D}^{\,*0\beta\dagger}\overline{D}^{\,*0}_{\beta}]+h_2[D^{*+\alpha\dagger}\overline{D}^{\,0}_{1\alpha}\overline{D}^{\,0\beta\dagger}_1D^{*+}_{\beta}\\
\label{10}&&-D^{*+\alpha\dagger}\overline{D}^{\,0\dagger}_{1\alpha}D^{*+\beta}\overline{D}^{\,0}_{1\beta}+D^{+\alpha\dagger}_1\overline{D}^{\,*0}_{\alpha}\overline{D}^{\,*0\beta\dagger}D^{+}_{1\beta}-D^{+\alpha\dagger}_1\overline{D}^{\,*0\dagger}_{\alpha}D^{+\beta}_1\overline{D}^{\,*0}_{\beta}]
\end{eqnarray}
A superscript $\dagger$ on a field represents its complex conjugate.
If the above interactions in Eq.(\ref{10}) is treated
nonperturbatively, there should be S-wave bound state which can be
identified with the Z(4430).

\begin{figure}
\begin{center}
\includegraphics*[width=8cm]{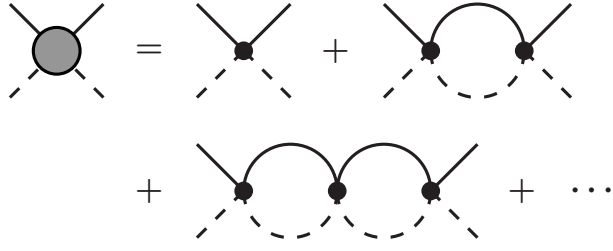}
\caption{\label{fig1}The Feynman diagrams for the transition
Z(4430)$\rightarrow$Z(4430) }
\end{center}
\end{figure}

We denote the transition amplitude for Z(4430)$\rightarrow$Z(4430)
by $i\mathcal{A}(E)$, it depends only on the total energy $E$ in the
center-of-mass frame, and Z(4430) corresponds to a pole of
$i\mathcal{A}(E)$. $i\mathcal{A}(E)$ can be calculated
nonperturbatively by summing the loop graphs in Fig.\ref{fig1}
\begin{equation}
\label{11}i\mathcal{A}(E)=\frac{4ih}{1-ihL(E)}
\end{equation}
where $h=h_1-3h_2$, and $L(E)$ is the amplitude for the propagation
of $D^{*}D_1$ between successive interaction vertex, it is given by
\begin{eqnarray}
\nonumber
L(E)&=&\int\frac{d^4q}{(2\pi)^4}\frac{i}{\frac{E}{2}-q_0-\Delta-\frac{\mathbf{q}^2}{2M_{D^{*}}}+i\epsilon}\,\frac{i}{\frac{E}{2}-q_0-\delta
m_T-\frac{\mathbf{q}^2}{2M_{D_1}}+i\epsilon}\\
\label{12}&=&-2M_{D^{*}D_1}i\int\frac{d^3\mathbf{q}}{(2\pi)^3}\frac{1}{\mathbf{q}^2-2M_{D^{*}D_1}(E-\Delta-\delta
m_T)-i\epsilon}
\end{eqnarray}
where $M_{D^{*}D_1}=M_{D^{*}}M_{D_1}/(M_{D^{*}}+M_{D_1})$ is the
reduced mass of the $D^{*}D_1$ system. From Eq.(\ref{12}), we see
that $L(E)$ has a linear ultraviolet divergence, which can be
removed by the renormalization of the coupling constant $h_1$ and
$h_2$. $L(E)$ is finite in dimensional regularization, which
corresponds to an explicit substraction of the linear divergence,
and we denote the renormalized $h$ as $h_R$. $L(E)$ in dimensional
regularization is given by
\begin{equation}
\label{13}L(E)=\frac{iM_{D^{*}D_1}}{2\pi}\sqrt{-2M_{D^{*}D_1}(E-\Delta-\delta
m_T)}
\end{equation}
Inserting the above expression for $L(E)$ into the amplitude
$i\mathcal{A}(E)$ in Eq.(\ref{11}), then the amplitude reduces to
\begin{equation}
\label{14}i\mathcal{A}(E)=\frac{ih_R}{1+{h_RM_{D^{*}D_1}}/{(2\pi)}\sqrt{-2M_{D^{*}D_1}(E-\Delta-\delta
m_T)}}
\end{equation}
If $h_{R}<0$, $\mathcal{A}(E)$ has a pole corresponding to Z(4430)
\begin{equation}
\label{15}E_{pole}=\Delta+\delta
m_T-\frac{2\pi^2}{h^2_RM^3_{D^{*}D_1}}
\end{equation}
The above energy is measured relative to twice the pseudoscalar mass
$M_D$, therefore the binding energy of Z(4430) is
\begin{equation}
\label{16}E_{Z,b}=M_{D^{*}}+M_{D_1}-(2M_D+\Delta+\delta
m_T-\frac{2\pi^2}{h^2_RM^3_{D^{*}D_1}})=\frac{2\pi^2}{h^2_RM^3_{D^{*}D_1}}
\end{equation}
If future experiments confirm the molecular state nature of Z(4430),
the binding energy of the bottom analog of Z(4430) can be predicted.
From the scaling of the coupling constants $h_1$ and $h_2$ with $M$
in Eq.(\ref{8}), we obtain $h_R(D)M_{D^{*}D_1}\sim
h_R(B)M_{B^{*}B_{1}}$. We denote the the bottom analog of Z(4430)
and its binding energy as $Z'$ and $E_{Z',b}$ respectively, then
\begin{equation}
\label{17}E_{Z',b}\sim E_{Z,b}\frac{M_{D^{*}D_1}}{M_{B^{*}B_1}}
\end{equation}
Since the $D^{*}D_{1}$ threshold is
$M_{D^{*+}}+M_{D^{0}_1}=4432.3\pm1.7$ MeV\cite{PDG}, and the Z(4430)
mass is $4433\pm4({\rm stat})\pm1({\rm syst})$ MeV, now we can not
determine the binding energy of Z(4430) because of the large
uncertainty. As an illustration, we assume $E_{Z,b}=1$ MeV for the
moment, then from Eq.(\ref{17}) we predict that the binding energy
of the bottom analog of Z(4430) is about 0.4 MeV and the mass
approximately is 11048.6 MeV. Our result for the mass of the bottom
analog of Z(4430) is larger than the quark model prediction
$(10730\pm100)$ MeV\cite{Cheung:2007wf} and the QCD sum rule
prediction $(10.74\pm0.12)$ GeV\cite{Lee:2007gs}. We note that the
effective theory prediction for the mass of $1^{++}$ $BB^{*}$
molecule state is larger than the quark model predictions as
well\cite{AlFiky:2005jd,Tornqvist:1993ng,Swanson:2006st}.

For the $PP_1$ sector, from Eq.(\ref{7}) we can obtain the relevant
contact interactions
\begin{equation}
\label{18}\mathcal{L}_{2,DD_1}=-h_1[P^{(Q)\dagger}_aP^{(Q)}_aP^{(\overline{Q})\beta\dagger}_{1b}P^{(\overline{Q})}_{1b\beta}+P^{(Q)\beta\dagger}_{1a}P^{(Q)}_{1a\beta}P^{(\overline{Q})\dagger}_bP^{(\overline{Q})}_b]
\end{equation}
This Lagrangian involves only one coupling constant $h_1$.
Performing similar calculations as for Z(4430), we find that only if
$h_{1R}>0$ there is a charged $DD_1$ molecule, and the binding
energy is ${2\pi^2}/{(h^2_{1R}M^3_{DD_1})}$, where $M_{DD_1}$ is the
reduced mass of $DD_1$ system. Therefore the existence of $D^{*}D_1$
molecule doesn't in general implies the existence of $DD_1$ bound
molecule.

The above discussions can be straightforwardly generalized to the
flavor SU(3) symmetry, and a nonet(octet and singlet) which Z(4430)
belongs to is predicted\cite{Liu:2007bf,Li:2007bh}. For the positive
charged $D^{*}_{s}D_1$(or $D_{s1}D^{*}$) molecule ${\rm Z^{+}_s}$,
its flavor wavefunction generally is $|{\rm
Z^{+}_s}\rangle=c_1|D^{*+}_{s}\overline{D}^{0}_1\rangle+c_2|D^{+}_{s1}\overline{D}^{*0}\rangle$
with $|c_1|^2+|c_2|^2=1$. The relevant four-boson contact
interactions are the $a=3,\,b=1$ terms in Eq.(\ref{9}), which is as
follows
\begin{eqnarray}
\nonumber{\cal L}_{2,{\rm
Z^{+}_s}}&=&h_1[D^{*+\alpha\dagger}_sD^{*+}_{s\alpha}\overline{D}^{\,0\beta\dagger}_1\overline{D}^{\,0}_{1\beta}+D^{+\alpha\dagger}_{s1}D^{+}_{s1\alpha}\overline{D}^{\,*0\beta\dagger}\overline{D}^{\,*0}_{\beta}]
+h_2[D^{*+\alpha\dagger}_s\overline{D}^{\,0}_{1\alpha}\overline{D}^{\,0\beta\dagger}_1D^{*+}_{s\beta}\\
\label{19}&&-D^{*+\alpha\dagger}_s\overline{D}^{\,0\dagger}_{1\alpha}D^{*+\beta}_s\overline{D}^{\,0}_{1\beta}+D^{+\alpha\dagger}_{s1}\overline{D}^{\,*0}_{\alpha}\overline{D}^{\,*0\beta\dagger}D^{+}_{s1\beta}-D^{+\alpha\dagger}_{s1}\overline{D}^{\,*0\dagger}_{\alpha}D^{+\beta}_{s1}\overline{D}^{\,*0}_{\beta}]
\end{eqnarray}
Therefore the binding energy of ${\rm Z^{+}_s}$ is
$2\pi^2/(h^2_RM^{3}_{D^{*}D_{s1}})$ under the SU(3) flavor symmetry.
It could decay into $D^{*}D^{*}_s\pi$, $D^{*}D^{*}K$ and
$K^{+}\psi'$, so the experimental search for these final states is
very interesting.

The assumption of Z(4430) as a weakly bound $D^{*}D_1$ molecule is
particularly predictive. The small binding energy, which is much
smaller other QCD scales such as $\Lambda_{QCD}$ and the pion mass
$m_{\pi}$ etc, implies that the molecule has universal
properties\cite{universal}. This can be further exploited through
factorization formulae to predict the production and decay
properties of Z(4430) just similar to the X(3872)
study\cite{factorization}. The relevant work is in
progress\cite{ding}.

\section{summary}

The proximity of Z(4430) mass to the $D^{*}D_1$ threshold favors a
$D^{*}D_{1}$ molecule state interpretation of Z(4430). In this work
we first discuss how to distinguish the molecule interpretation from
the tetraquark picture, and we find that experimentally measuring
whether Z(4430) dominantly decays into ${D^{*}D^{*}\pi}$ or the two
body modes ${DD^{*}}$, ${D^{*}D^{*}}$, {$J/\psi\pi$, $J/\psi\rho$,
$\eta_c(1{\rm S})\rho$ and $\psi(2{\rm S})\pi$} is very important
for understanding the structure of Z(4430). If Z(4430) is a
tetraquark state, another state with mass about 4344 MeV should be
found in the $\psi'\pi^{+}$ final state in addition.

We have studied Z(4430) as a $D^{*}D_1$ molecule from the effective
field theory, and an effective Lagrangian with four-boson contact
interactions is constructed to describe this system. We find that if
Z(4430) is a $D^{*}D_1$ molecule state, there should be a $B^{*}B_1$
molecule with mass about 11048.6 MeV. The existence of the
$D^{*}D_1$ bound state doesn't in general implies a bound state in
the $DD_1$ system.

\section*{ACKNOWLEDGEMENTS}
\indent  I am grateful to Prof. Mu-Lin Yan, Prof. Dao-Neng Gao and
Dr. Jie-Jie Zhu for very helpful and stimulating discussions.


\end{document}